\documentclass{article}
\usepackage{amsmath,enumerate}
\usepackage{amssymb,amsthm}
\usepackage[pdftex]{graphicx}
\usepackage{subfigure,epsfig, epstopdf}
\usepackage{soul}
\usepackage{natbib}
\usepackage{float}

\newfloat{algorithm}{tbph}{alg}
\floatname{algorithm}{Algorithm}

\setcitestyle{authoryear,open={(},close={)}}

\renewcommand{\d}{\,\mathrm{d}}
\newcommand{\sF}{\mathcal{F}}
\newcommand{\sG}{\mathcal{G}}

\newcommand{\sL}{\mathcal{L}}
\newcommand{\bw}{\mathbf{w}}
\newcommand{\bx}{\mathbf{x}}

\newcommand{\sE}{\mathcal{E}}

\newtheorem{theorem}{Theorem}

\title{Efficient recycled algorithms for quantitative trait models on phylogenies}
\author{Gordon Hiscott\textsuperscript{1} \and Colin Fox\textsuperscript{2} \and Matthew Parry\textsuperscript{1} \and 
 David Bryant\textsuperscript{1}}
\date{\small \textsuperscript{1}Department of Mathematics and Statistics,  \textsuperscript{2}Department of Physics, \\University of Otago, Dunedin, New Zealand \\ 
}
\begin{document}

\maketitle

\begin{abstract}
We present an efficient and flexible method for computing likelihoods of phenotypic traits on a phylogeny. The method does not resort to Monte-Carlo computation but instead blends Felsenstein's discrete character pruning algorithm with methods for numerical quadrature. It is not limited to Gaussian models and adapts readily to model uncertainty in the observed trait values. We demonstrate the framework by developing efficient algorithms for likelihood calculation and ancestral state reconstruction under Wright's threshold model, applying our methods to a dataset of trait data for extrafloral nectaries (EFNs) across a phylogeny of 839 Labales species.
\end{abstract}


\newpage

\section*{Introduction}

Models for the evolution of continuous traits on a phylogeny are an essential tool of evolutionary genomics. These models are used to describe a huge range of biological phenomena, including morphological characteristics \citep{Felsenstein2002a,HarmonETAL2010a,Ronquist2004a,Stevens1991a}, expression levels \citep{KhaitovichETAL2005a,KhaitovichETAL2006a},  geography \citep{LemeyETAL2010a}, and gene frequency dynamics \citep{Cavalli-SforzaETAL1967a,Felsenstein1981a,SirenETAL2011a}. Model-based statistical techniques can detect or test for association between different traits, reconstruct ancestral states, and infer subtle features of the evolutionary process that generated them. 

The usefulness of continuous trait models is contingent on our ability to compute with them. Early work of Felsenstein \citep{Felsenstein1968a,Felsenstein1973a} demonstrated that if traits are evolving according to Brownian motion then we can compute likelihoods quickly and (up to numerical precision) exactly. Felsenstein's method extends directly to other Gaussian processes, notably the Ornstein-Uhlenbeck (OU) process \citep{Felsenstein1988a,Hansen1997a,Lande1976a}. These methods break down for more complex models, in which case researchers have typically resorted to Monte Carlo strategies (e.g. \citep{LandisETAL2013a,Ronquist2004a}). 

Computing the probability of a qualitative character is essentially a numerical integration (quadrature) problem. For most models, if we know the value of the trait at each ancestral node in the phylogeny we can quickly compute the various transition probabilities. Since we do not usually know these ancestral trait values we integrate them out. This is a multi-dimensional integration problem with one dimension for each ancestral node (or two dimensions for each node if we are modelling covarying traits).

Methods for estimating or approximating integrals are usually judged by their {\em rate of convergence}: how quickly the error of approximation decreases as the amount of work (function evaluations) increases. Consider the problem of computing a one-dimensional integral
\begin{equation} \int_0^1 \! f(x) \d x\end{equation} 
where $f$ is a `nice' function with continuous and bounded derivatives. Simpson's rule, a simple textbook method reviewed below, can be shown to have an $O(N^{-4})$ rate of convergence, meaning that, asymptotically in $N$, evaluating 10 times more points reduces the error by a factor of $10^{4}$. In contrast, a standard Monte Carlo method  has a rate of convergence of $O(N^{-\frac{1}{2}})$, meaning that evaluating 10 times more points will only reduced the error by a factor of around 3 For this reason, numerical analysis texts often refer to Monte Carlo approaches as `methods of last resort.'

Despite this apparently lacklustre performance guarantee, Monte-Carlo methods have revolutionised phylogenetics in general and the analysis of quantitative characters in particular. The reason is their partial immunity to the curse of dimensionality. Methods like Simpson's rule are not practical for a high number of dimensions as the asymptotic convergence rate, quoted above, is only achieved for an infeasibly large number of function evaluations $N$. The effective convergence rate for small $N$ can be very poor, and typically worse than Monte-Carlo. In contrast, there are Monte Carlo approaches which achieve close to $O(N^{-\frac{1}{2}})$ convergence irrespective of dimension. This has been critical when computing with complex evolutionary models with as many dimensions as there are nodes in the phylogeny. 

The main contribution of our paper is to demonstrate how to efficiently and accurately compute likelihoods on a phylogeny using a sequence of one-dimensional integrations. We obtain a fast algorithm with convergence guarantees that far exceed what can be obtained by Monte Carlo integration. Our approach combines two standard tools: classical numerical integrators and Felsenstein's pruning algorithm for {\em discrete characters} \citep{Felsenstein1981b}. Indeed, the only real difference between our approach and Felsenstein's discrete character algorithm is that we use numerical integration techniques to integrate states at ancestral nodes, instead of just carrying out a summation. 

The running time of the algorithm is  $O(N^2n)$, where $N$ is the number of points used in the numerical integration at each node and $n$ is the number of taxa (leaves) in the tree. Using Simpson's method, we obtain a convergence rate of $O(nN^{-4})$, meaning that if we increase $N$ by a factor of $10$ we will obtain an estimate which is accurate to four more decimal places.

To illustrate the application of our general framework, we develop an efficient algorithm for computing the likelihood of a tree under the threshold model of Sewell Wright and Felsenstein \citep{Felsenstein2005a,Felsenstein2012a,Wright1934a}. We also show how to infer marginal trait densities at ancestral nodes. We have implemented these algorithms and used them to study evolution of extrafloral nectaries on an 839-taxon phylogeny \cite{MarazziETAL2012a}.

The combination of numerical integrators and the pruning algorithm opens up a large range of potential models and approaches which we have only just begun to explore. In the discussion, we briefly review developments in numerical integration techniques that could well be brought to bear on these problems, and a few suggestions of directions and problems which can now be addressed.

\section*{Material and Methods}

\subsubsection*{Models for continuous trait evolution}

Phylogenetic models for continuous trait evolution, like those for discrete traits, are specified by the density of trait values at the root and the transition densities along the branches. We use $f(x_r|\theta_r)$ to denote the density for the trait value at the root, where $\theta_r$ is a set of relevant model parameters. We use 
$f(x_i|x_j,\theta_i)$ to denote the transitional density for the value at node $i$, conditional on the trait value at its parent node $j$. Here, $\theta_i$ represents a bundle of parameters related to node $i$ such as branch length, population size, and mutation rate. All of these parameters could vary throughout the tree. 

To see how the model works, consider how continuous traits might be simulated. 
A state $X_r$ is sampled from  the root density $f(X_r|\theta_r)$. We now proceed through the phylogeny from the root to the tips, each time visiting a node only after its parent has already been visited. For each node $i$, we generate the value at that node from the density $f(X_i|x_j,\theta_v)$, where $x_j$ is the simulated trait value at node $j$, the parent of node $i$. In this way, we will eventually generate trait values for the tips. 

We use $X_1,\ldots,X_n$ to denote the random trait values at the tips and $X_{n+1},\ldots,X_{2n-1}$ to denote the random trait values at the internal nodes, ordered so that children come before parents. Hence $X_{2n-1}$ is the state assigned to the root. Let
\begin{equation} \sE(T) = \{(i,j):\mbox{node $i$ is a child node $j$}\}\end{equation} 
denote the set of branches in the tree. 
The joint density for all trait values, observed and ancestral, is given by multiplying the root density with all of the transition densities
\begin{equation} f(x_1,\ldots,x_n,x_{n+1},\ldots,x_{2n-1}|\theta) = f(x_{2n-1}|\theta) \! \! \! \prod_{(i,j) \in \sE(T)} \! \! \! f(x_i|x_j,\theta_i).\end{equation} 
The probability of the observed trait values $x_1,\ldots,x_n$ is now determined by {\em integrating} out all of the ancestral trait values:
\begin{equation} \label{eq:allTips}
\sL(T) = f(x_1,\ldots,x_n|\theta) = \int \int \cdots \int f(x_{2n-1}|\theta_r) \! \! \!  \prod_{(i,j) \in \sE(T)}  \! \! \!  f(x_i|x_j,\theta_i)\, \d x_{n+1},\ldots, \d x_{2n-1}.
\end{equation}
In these integrals, the bounds of integration will vary according to the model.

The oldest, and most widely used, continuous trait models assume that traits (or transformed gene frequencies) evolve like Brownian motion \citep{Cavalli-SforzaETAL1967a,Felsenstein1973a}. For these models, the root density $f(x_r|\theta)$ is Gaussian (normal) with mean $0$ and unknown variance $\sigma_r^2$. The transition densities $f(x_i|x_j,\theta_v)$ are also Gaussian, with mean $x_j$ (the trait value of the parent) and standard deviation proportional to branch length. Note that there are identifiability issues which arise with the inference of the root position under this model, necessitating a few tweaks in practice. 

It can be shown that when the root density and transitional densities are all Gaussian,  the joint density \eqref{eq:allTips} is multivariate Gaussian. Furthermore, the covariance matrix for this density has a special structure which methods such as the pruning technique of \citet{Felsenstein1973a} exploit. This technique continues to work when Brownian motion is replaced by an OU process  \citep{Felsenstein1988a,Hansen1997a,Lande1976a}.

\citet{LandisETAL2013a} discuss a class of continuous trait models which are based on {\em L\'{e}vy processes} and include jumps. At particular times, as governed by a Poisson process, the trait value jumps to a value drawn from a given density. Examples include a {\em compound Poisson process} with Gaussian jumps and a {\em variance Gamma} model with Gamma distributed jumps. Both of these processes have analytical transition probabilities in some special cases. 

\citet{LepageETAL2006a} use the Cox-Ingersoll-Ross (CIR) process to model rate variation across a phylogeny. Like the OU process (but unlike Brownian motion), the CIR process is ergodic. It has a stationary Gamma density which can be used for the root density. The transition density is a particular non-central Chi-squared density and the process only assumes positive values.

  \citet{KutsukakeETAL2013a} examine a family of compound Poisson models, focusing particularly on a model where the trait values make exponentially distributed jumps upwards or downwards. In the case that the rates of upward and downward jumps are the same, the model reduces to the variance Gamma model of \citet{LandisETAL2013a} and has an analytical probability density function.

 \citet{SirenETAL2011a} propose a simple and elegant model for gene frequencies whereby the root value is drawn from a Beta distribution and each transitional density is Beta with appropriately chosen parameters. 

Trait values at the tips are not always observed directly. A simple, but important, example of this is the threshold model of \citet{Wright1934a}, explored by \citet{Felsenstein2005a}. Under this model, the trait value itself is censored and we only observe whether or not the value is positive or negative. A similar complication arises when dealing with gene frequency data as we typically do not observe the actual gene frequency but instead a binomially distributed sample based on that frequency \citep{SirenETAL2011a}. 

If the trait values at the tip are not directly observed we integrate over these values as well. Let $\pi(z_i|x_i,\theta_i)$ denote the probability of observing  $z_i$ given the trait value $x_i$. The marginalised likelihood is then 
\begin{equation} \sL(T|z_1,\ldots,z_n) = \int \int \cdots \int f(x_r|\theta)  \! \! \! \prod_{(i,j) \in \sE(T)}  \! \! \!  f(x_i|x_j,\theta_v)    \! \prod_{i=1}^n  \pi(z_i|x_i,\theta_i)\, \d x_{1},\ldots, \d x_{2n-1}.\end{equation}

\subsubsection*{Numerical integration}

Analytical integration can be difficult or impossible. For the most part, it is unusual for an integral to have an analytical solution, and there is no general method for finding it when it does exist. In contrast,  {\em numerical integration} techniques (also known as {\em numerical quadrature}) are remarkably effective and are often easy to implement. A numerical integration method computes an approximation of the integral from function values at a finite number of points. Hence we can obtain approximate integrals of functions even when we don't have an equation for the function itself. See \cite{CheneyETAL2012a} for an introduction to numerical integration, and \cite{DahlquistETAL2008a} and \cite{DavisETAL2007a} for more comprehensive technical surveys.

The idea behind most numerical integration techniques is to approximate the target function using a function which is easy to integrate. In this paper we will restrict our attention to {\em Simpson's method} which approximates the original function using  piecewise quadratic functions. To approximate an integral $\int_a^b f(x)  \d x$ we first determine $N+1$ equally spaced points 
\begin{equation} x_0 = a,\,\, \,\,x_1 =  a + \frac{b-a}{N},  \,\,  \,\,  x_2 = a + 2\frac{b-a}{N} , \ldots, x_k = a + k\frac{b-a}{N},\ldots,\,\,x_N = b.\end{equation} 
We now divide the integration into $N/2$ intervals 
\begin{equation} \int_a^b f(x) \, \d x = \sum_{\ell = 1}^{N/2} \,\,\,\,\int \displaylimits_{x_{2\ell - 2}}^{x_{2 \ell}} f(x) \, \d x.\end{equation} 
Within each interval $[x_{2\ell - 2},x_{2 \ell}]$, there is a unique quadratic function which equals $f(x)$ at each the three points $x = x_{2 \ell - 2}$, $x = x_{2 \ell -1}$ and $x = x_{2 \ell}$.  The integral of this quadratic on the interval $[x_{2\ell-2},x_{2 \ell}]$
is $\frac{(b-a)}{3N} \left( f(x_{2\ell-2}) + 4f(x_{2\ell-1}) + f(x_{2\ell}) \right)$. Summing over $\ell$, we obtain the approximation
\begin{equation} \int_a^b f(x) \, \d x \approx \sum_{\ell=1}^{N/2} \frac{(b-a)}{3N} \left( f(x_{2\ell-2}) + 4f(x_{2\ell-1}) + f(x_{2\ell}) \right).\end{equation} 
With a little rearrangement, the approximation can be written in the form 
\begin{equation}
\int_a^b f(x) \, \d x \approx \frac{(b-a)}{N} \sum_{k=0}^N w_k f(x_k) \label{eq:wvec}
\end{equation}
where $w_k = 4/3$ when $k$ is odd and $w_k = 2/3$ when $k$ is even, with the exception of 
$w_0$ and $w_N$ which both equal $1/3$. Simpson's method is embarrassingly easy to implement and has a convergence rate of $O(N^{-4})$. Increasing the number of intervals by a factor of $10$ decreases the error by a factor of $10^{-4}$.  See \cite{DahlquistETAL2008a} and \cite{DavisETAL2007a} for further details.

It should be remembered, however, that the convergence rate is still only an asymptotic bound, and gives no guarantees on how well the method performs for a specific function and choice of $N$. Simpson's method, for example, can perform quite poorly when the function being integrated has rapid changes or soft peaks. We observed this behaviour when implementing threshold models, as described below. Our response was to better tailor the integration method 
for the functions appearing. We noted that the numerical integrations we carried out all had the form 
\begin{equation} \int_a^b e^{-\frac{(x-\mu)^2}{2\sigma^2}} f(x) \, \d x\end{equation} 
where $\mu$ and $\sigma$ varied. Using the same general approach as Simpson's rule, we approximated $f(x)$, rather than the whole function $e^{-\frac{(x-\mu)^2}{2\sigma^2}} f(x)$, by a piecewise quadratic function $p(x)$. We could then use standard techniques and tools to evaluate $\int_a^b e^{-\frac{(x-\mu)^2}{2\sigma^2}} p(x) \, \d x$ numerically. The resulting integration formula, which we call the {\em Gaussian kernel method}, gives a significant improvement in numerical accuracy.

A further complication is that, in models of continuous traits, the trait value  often ranges over the whole real line, or at least over the set of positive reals. Hence, we need to approximate integrals of the form 
\begin{equation} \int_{-\infty}^\infty f(x) \, \d x \mbox{ or } \int_0^\infty f(x) \, \d x\end{equation} 
though the methods discussed above only apply to integrals on finite intervals. We truncate these integrals, determining values $U$ and $L$ such that the difference 
\begin{equation} \int_{-\infty}^\infty f(x) \, \d x - \int_{L}^U f(x) \, \d x\end{equation} 
between the full integral $\int_{-\infty}^\infty f(x) \, \d x$ and the truncated integral $\int_{L}^U f(x) \, \d x$
can be bounded analytically. Other strategies are possible; see \cite{DahlquistETAL2008a} for a comprehensive review.

\subsubsection*{A pruning algorithm for integrating continuous traits}

Felsenstein has developed pruning algorithms for both continuous and discrete characters  \citep{Felsenstein1981a,Felsenstein1981b}. His algorithm for continuous characters works only for Gaussian processes. Our approach is to take his algorithm for {\em discrete characters} and adapt  it to continuous characters.

The (discrete character) pruning algorithm is an application of dynamic programming. For each node $i$, and each state $x$, we compute the probability of observing the states for all tips which are descendants of node $i$, conditional on node $i$ having ancestral state $x$. This probability is called the {\em partial likelihood} at node $i$ given state $x$. Our algorithm follows the same scheme, with one major difference. Since traits are continuous, we cannot store all possible partial likelihoods. Instead, we store likelihoods for a finite set of values and plug these values into a numerical integration routine.

Let $i$ be the index of a node in the tree not equal to the root, let node $j$ be its parent node. We define the {\em partial likelihood}, $\sF_i(x)$ to be the likelihood for 
 the observed trait values at the tips which are descendants of node $i$, conditional on the parent node $j$ having trait value $x$. If node $i$ is a tip with observed trait value $x_i$ we have 
\begin{equation}
\sF_i(x) = f(x_i|x,\theta_i) \label{eq:leaf1}
\end{equation}
recalling that $f(x_i |x,\theta_i)$ is the density for the value of the trait at node $i$ conditional on the value of the trait for its parent. 
More generally, we may only observe some value $z_i$ for which we have the conditional probability $\pi(z_i|x_i,\theta_i)$ conditional on the trait value $x_i$. In this case the partial likelihood is given by
\begin{equation}
\sF_i(x) = \int f(\tilde{x}|x,\theta_i) \pi(z_i|\tilde{x}) \, \d \tilde{x}. \label{eq:leaf3}
\end{equation}

Suppose node $i$ is not the root and that it  has two children $u,v$. Since trait evolution is conditionally independent on disjoint subtrees, we obtain the recursive formula
\begin{equation}
\sF_i(x) = \int f(\tilde{x} | x,\theta_i) \sF_{u}(\tilde{x}) \sF_{v}(\tilde{x})\, d\tilde{x}. \label{eq:recurse}
\end{equation}
Finally, suppose that node $i$ is the root and has two children $u,v$. We evaluate the complete tree likelihood using the density of the trait value at the root, 
\begin{equation}
\sL(T) = \int f(x|\theta_r) \sF_{u}(x) \sF_{v}(x)\, \d x. \label{eq:root}
\end{equation}
The bounds of integration in \eqref{eq:leaf3}---\eqref{eq:root} will vary according to the model.

We use numerical integration techniques to approximate  \eqref{eq:leaf3}---\eqref{eq:root}  and dynamic programming to avoid an exponential explosion in the computation time. Let $N$ denote the number of function evaluations for each node. In practice, this might vary over the tree, but for simplicity we assume that it is constant. For each node $i$, we select $N+1$ trait values
\begin{equation} X_i[0] < X_i[1] < \cdots < X_i[N].\end{equation} 
How we do this will depend on the trait model and the numerical integration technique. If, for example, the trait values vary between $a$ and $b$ and we are applying  Simpson's method with $N$ intervals we would use 
$X_i[k] = a+\frac{b-a}{N} k$ for $k=0,1,2,\ldots,N$.

We traverse the tree starting at the tips and working towards the root. For each non-root node $i$ and $k=0,1,\ldots,N$ we compute and store an approximation $F_i[k]$ of $\sF_i(X_j[k])$, where node $j$ is the parent of node $i$.  Note that this is an approximation of $\sF_i(X_j[k])$ rather than of $\sF_i(X_i[k])$ since $\sF_i(x)$ is the partial likelihood conditional on the trait value for the {\em parent} of node $i$. The value approximation $F_v[i]$ is computed by applying the numerical integration method to the appropriate integral  \eqref{eq:leaf3}---\eqref{eq:root}, where we replace function evaluations with approximations previously computed. See below for a worked example of this general approach.

The numerical integration methods we use run in time linear in the number of points being evaluated. Hence if $n$ is the number of tips in the tree, the algorithm will run in time $O(nN^2)$.  For the integration techniques described above, the convergence rate (in $N$) for the likelihood on the entire tree had the same order as the convergence rate for the individual one-dimensional integrations (see below for a formal proof of a specific model). We have therefore avoided the computational blow-out typically associated with such high-dimensional integrations, and achieve this without sacrificing accuracy.

\subsubsection*{Posterior densities for ancestral states}

The algorithms we have described compute the joint density of the states at the tips, given the tree, the branch lengths, and other parameters. As with discrete traits, the algorithms can be modified to infer ancestral states for internal nodes in the tree. Here we show how to carry out reconstruction of the marginal posterior density of a state at a particular node. The differences between marginal and joint reconstructions are reviewed in \cite[pg 121]{Yang2006a}.

First consider marginal reconstruction of ancestral states at the root. Let $u$ and $v$ be the children of the root. The product  $\sF_{u}(x) \sF_{v}(x)$ equals the probability of the observed character conditional on the tree, branch lengths, parameters and a state of $x$ at the root. The marginal probability of $x$, ignoring the data, is given by the root density $f(x|\theta_r)$. Integrating the product of $\sF_{u}(x) \sF_{v}(x)$ and $f(x|\theta_r)$ gives the likelihood $\sL(T)$, as in \eqref{eq:root}. Plugging these into Bayes' rule, we obtain the {\em posterior density} of the state at the root:
\begin{equation} f(x_r|z_1,\ldots,z_n) = \frac{\sF_{u}(x_r) \sF_{v}(x_r) f(x_r|\theta_r)}{\sL(T)}.\end{equation} 
With general time reversible models used in phylogenetics, the posterior distributions at other nodes can be found by changing the root of the tree. Unfortunately the same trick does not work for many qualitative trait models, including the threshold model we study here. Furthermore, recomputing likelihoods for each possible root entails a large amount of unneccessary computation. 

Instead we derive a second recursion, this one starting at the root and working towards the tips. A similar trick is used to compute derivatives of the likelihood function in \cite{FelsensteinETAL1996a}. For a node $i$ and state $x$ we let $\sG_i(x)$ denote the likelihood for the trait values at tips which are {\em not} descendants of node $i$, conditional on node $i$ having trait value $x$. If node $i$ is the root $r$, then $\sG_r(x)$ is $1$ for all $x$. 

Let node $i$ be any node apart from the root, let node $j$ be its parent and let node $u$ be the other child of $j$ (that is, the sibling of node $i$). We let $\tilde{x}$ denote the trait value at node $j$. Then $\sG_i(x)$ can be written 
\begin{equation}
\sG_i(x) = \int f(\tilde{x} |x,\theta_i) \sG_j(\tilde{x}) \sF_u(\tilde{x}) \, d \tilde{x}. \label{eq:postGeneral}
\end{equation}
This integral can be evaluated using the same numerical integrators used when computing likelihoods. Note that $f(\tilde{x} |x,\theta_i)$ is the conditional density of the {\em parent} state given the child state, which is the reverse of the transition densities used to formulate the model. How this is computed will depend on the model and its properties; see below for an implementation of this calculation in the threshold model. 

Once $\sG_i(x)$ has been computed for all nodes, the actual (marginal) posterior densities are computed from Bayes' rule. Letting $u,v$ be the children of node $i$,
\begin{equation}
f(x_i|z_1,\ldots,z_n) = \frac{\sG_i(x_i) \sF_u(x_i) \sF_v(x_i) f(x_i)}{\sL(T)}. \label{eq:posteq}
\end{equation}

\subsubsection*{Case study: threshold models}

In this section we show how the general framework can be applied to  the threshold model of \cite{Wright1934a} and \cite{Felsenstein2005a,Felsenstein2012a}. 
Each trait is modelled by a continuously varying {\em liability} which evolves along branches according to a Brownian motion process. While the underlying liability is continuous, the observed data is discrete: at each tip we observe only whether the  liability is above or below some threshold.

We will use standard notation for Gaussian densities. Let $\phi(x|\mu,\sigma^2)$ denote the density of a Gaussian random variable $x$ with mean $\mu$ and variance $\sigma^2$; let
\begin{equation} \Phi(y|\mu,\sigma^2) = \int_{-\infty}^y \phi(x|\mu,\sigma^2)\end{equation} 
denote its cumulative density function, with inverse $\Phi^{-1}(\alpha | \mu,\sigma^2)$.

Let $X_1,\ldots,X_{2n-1}$ denote the (unobserved) liability values at the $n$ tips and $n-1$ internal nodes. As above we assume that the $i < j$ whenever node $i$ is a child of node $j$, so that the root has index $2n-1$. 

The liability value at the root has a Gaussian density with mean $\mu_r$ and variance $\sigma_r^2$:
\begin{equation} f(x_{2n-1}|\theta_r) = \phi(x_{2n-1}|\mu_r,\sigma_r^2).\end{equation} 
Consider any non-root node $i$ and let $j$ be the index of its parent. Let $t_i$ denote the length of the branch connecting nodes $i$ and $j$. Then $X_i$ has a Gaussian density with mean $x_j$ and variance $\sigma^2 t_v$:
\begin{equation} f(x_{i}|x_j,\theta_i) = \phi(x_i|x_j,\sigma^2 t_i).\end{equation} 
Following \cite{Felsenstein2005a}, we assume thresholds for the tips are all set at zero. We observe  $1$ if the liability is positive, $0$ if the liability is negative, and $?$ if data is missing.  We can include the threshold step into our earlier framework by defining
\begin{equation} \pi(z_i|x_i) = \begin{cases} 1 & \mbox{ if $z_i = 1$ and $x_i > 0$,  or $z_i = 0$ and $x_i \leq 0$, or $z_i=?$} \\
0 & \mbox{ otherwise.} \end{cases}\end{equation} 
The likelihood function for observed discrete values $z_1,\ldots,z_n$ is then given by integrating over liability values for all nodes on the tree:
\begin{equation} \sL(T|z_1,\ldots,z_n) = \int \displaylimits_{-\infty}^{\infty} \! \cdots \! \int \displaylimits_{-\infty}^{\infty}    \phi(x_{2n-1}|\mu_r,\sigma_r^2) \prod_{(i,j)} \! \! \phi(x_i|x_j,\sigma^2 t_i) \prod_{i=1}^n \pi(z_i|x_i) \, \d x_{1} \ldots  \d x_{2n-1}.\end{equation}

The first step towards computing $\sL(T|z_1,\ldots,z_n)$ is to bound the domain of integration so that we can apply Simpson's method. Ideally, we would like these bounds to be as tight as possible, for improved efficiency. For the moment we will just outline a general procedure which can be adapted to a wide range of evolutionary models.

The {\em marginal (prior) density} of a single liability or trait value at a single node is the density for that liability value marginalizing over all other values and data. With the threshold model, the marginal density for the liability at node $i$ is Gaussian with mean $\mu_r$ (like the root) and variance $v_i$ equal to the sum of the variance at the root and the transition variances on the path from the root to node $i$. If $P_i$ is the set of nodes from the root to node $i$, then
\begin{equation}
v_i = \sigma_r^2 + \sigma^2 \sum_{j \in P_i} t_j.
\label{eq:marginalv}
\end{equation}

The goal is to constrain the error introduced by truncating the integrals with infinite domain. Let $\epsilon$ be the desired bound on this truncation error. Recall that the number of internal nodes in the tree is $n-1$. Define
\begin{equation} L_i = \Phi^{-1}\left(\frac{\epsilon}{2(n-1)}\Big|\mu_r,v_i\right)\end{equation} 
and
\begin{equation} U_i = \Phi^{-1}\left(\frac{\epsilon}{2(n-1)}\Big|\mu_r,v_i\right)\end{equation} 
so that the probability $X_i$ lies outside the interval $[L_i,U_i]$ is at most $\epsilon/(n-1)$. By the inclusion-exclusion principle, the joint probability $X_i \not \in [L_i,U_i]$ for {\em any} internal node $i$ is at most $\epsilon$. We use this fact to bound the contribution of the regions outside these bounds. 
\begin{align} \int \displaylimits_{-\infty}^{\infty} & \! \cdots \! \int \displaylimits_{-\infty}^{\infty}     f(x_{2n-1}|\mu_r,\sigma_r^2) \prod_{(u,v)} \! \! f(x_v|x_u,\theta_v)  \prod_{i=1}^n \pi(z_i|x_i) \, \d x_{1} \ldots \d x_{2n-1}  \\ 
& \qquad - \int \displaylimits_{a_{2n-1}}^{b_{2n-1}} \! \! \cdots \!  \int \displaylimits_{a_{n+1}}^{b_{n+1}}  \int \displaylimits_{-\infty}^{\infty}     \! \cdots \! \int \displaylimits_{-\infty}^{\infty}     f(x_{2n-1}|\mu_r,\sigma_r^2) \prod_{(u,v)} \! \! f(x_v|x_u,\theta_v) \prod_{i=1}^n \pi(z_i|x_i) \, \d x_{1} \ldots \d x_{2n-1} \\
& \leq  \int \displaylimits_{-\infty}^{\infty}  \! \cdots \! \int \displaylimits_{-\infty}^{\infty}     f(x_{2n-1}|\mu_r,\sigma_r^2) \prod_{(u,v)} \! \! f(x_v|x_u,\theta_v) \, \d x_{1} \ldots \d x_{2n-1}  \\ 
& \qquad - \int \displaylimits_{a_{2n-1}}^{b_{2n-1}} \! \! \cdots \!  \int \displaylimits_{a_{n+1}}^{b_{n+1}}  \int \displaylimits_{-\infty}^{\infty}     \! \cdots \! \int \displaylimits_{-\infty}^{\infty}     f(x_{2n-1}|\mu_r,\sigma_r^2) \prod_{(u,v)} \! \! f(x_v|x_u,\theta_v) \, \d x_{1} \ldots \d x_{2n-1}   \\
&\leq P\Big(X_{n+1} \not \in [L_{n+1},U_{n+1}] \mbox{ or }  X_{n+2} \not \in [L_{n+2},U_{n+2}] \mbox{ or } \cdots \mbox{ or } X_{2n-1} \not \in [L_{2n-1},U_{2n-1}] \Big) \\
& < \epsilon.
\end{align}
We therefore compute values $L_i,U_i$ for $n+1 \leq i \leq 2n-1$ and use these bounds when carrying out integration at the internal nodes. We define
\begin{equation} X_i[k] = L_i + \frac{U_i - L_i}{N} k\end{equation} 
for $k=0,1,\ldots,N$ for each internal node $i$.

The next step is to use dynamic programming and numerical integration to compute the approximate likelihood. 
Let node $i$ be a tip of the tree, let node $j$ be its parent and let $z_i$ be the binary trait value at this tip. 
For each $k=0,1,\ldots,N$  we use standard error functions to compute
\begin{eqnarray}
 F_i[k] &=&  \sF_i(X_j[k]) \\
 & = & \begin{cases} 
\int  \displaylimits_{0}^{\infty}   \phi(\tilde{x}|X_j[k],\sigma^2 t_i)  \, \d \tilde{x}  & \mbox{ if $z_i = 1$}\\
\int  \displaylimits_{-\infty}^{0}   \phi(\tilde{x}|X_j[k],\sigma^2 t_i)  \, \d \tilde{x}  & \mbox{ if $z_i = 0$}\\
1 & \mbox{ if $z_i = ?$.} \end{cases}
\end{eqnarray}
Here $\phi(x|\mu,\sigma^2)$ is the density of a Gaussian with mean $\mu$ and variance $\sigma^2$. 

Now suppose that node $i$ is an internal node with parent node $j$ and children $u$ and $v$. Applying Simpson's rule to the bounds $L_i,U_i$ to \eqref{eq:recurse} we have for each $k=0,1,\ldots,N$:
\begin{eqnarray}
F_i[k] & = & \frac{U_i-L_i}{N} \sum \displaylimits_{\ell = 0}^N w_\ell \phi(X_i[\ell]| X_j[k],\sigma^2 t_i) F_u[\ell] F_v[\ell] \label{eq:Frecurse}\\
& \approx & \sF_i(X_j[k]).
\end{eqnarray}
Suppose node $i$ is the root, and $u,v$ are its children. Applying Simpson's rule to \eqref{eq:root} gives
\begin{eqnarray} L &\leftarrow & \frac{U_{2n-1} - L_{n-1}}{N} \sum_{\ell = 0}^N w_\ell \phi(X_i[\ell]| \mu_r,\sigma_r^2) F_u[\ell] F_v[\ell] \\
& \approx & \sL(T|z_1,\ldots,z_n,\theta).
\end{eqnarray}

Pseudo-code for the algorithm appears in Algorithm~\ref{algo:OneChar}. Regarding efficiency and convergence we have:

\begin{theorem}
Algorithm~\ref{algo:OneChar} runs in $O(nN^2)$ time and approximates $L(T)$ with $O(nN^{-4})$ error.
\end{theorem}
\noindent {\em Proof}\\
The running time follows from the fact that for each of the $O(n)$ nodes in the tree we carry out $O(N)$ applications of Simpson's method. 

Simpson's rule has $O(N^{-4})$ convergence on functions with bounded fourth derivatives \citep{DahlquistETAL2008a}. The root density and each of the transition densities are Gaussians, so have individually have bounded fourth derivatives. For each node $i$, let $n_i$ denote the number of tips which are descendents of the node. Using induction on \eqref{eq:recurse}, we see that for all nodes $i$, the fourth derivate of $\sF_i(x)$ is $O(n_i)$. 

Letting $\epsilon = nN^{-4}$ we have from above that the error introduced by replacing the infinite domain integrals with integrals on $[L_i,U_i]$ introduces at most $nN^{-4}$ error. Using a second induction proof on \eqref{eq:recurse} and \eqref{eq:Frecurse} together with the bound on fourth derivatives we have that  $|\sF_i(X_j[k]) - F_i[k]|$ is at most $O(n_i N^{-4})$ for all nodes $i$, where node $j$ is the parent of node $i$.  In this way we obtain  error bound of $O(n_{2n-1}N^{-4}) = O(nN^{-4})$ on the approximation of $\sL(T | z_1,\ldots,z_n,\theta)$. \hfill $\Box$\\

\begin{algorithm}[htb]
\begin{sffamily}
\begin{tabbing}
\=XXX\=XXX\=XXX\=XXX\=XXX\=XXX\= \kill
{\bf Algorithm 1}: Compute probability of a threshold character.\\ ~\\
\> \> {\bf Input:} \\
\> \> \> $N$: Number of intervals in numerical integration.\\
\> \> \> $t_1,\ldots,t_{2n-2}$: branch lengths in tree.\\
\> \> \> $\mu_r,\sigma^2_r$: mean and variance of root density\\
\> \> \> $\sigma^2$: variance of transition densities (per unit branch length)\\
\> \> \> $z_1,\ldots,z_n$ observed character ($z_i \in \{+1,0,?\}$)\\
\> \> {\bf Output:}\\
\> \> \> Probability $L$ of observed character under the threshold model.\\
\\
\> \> Construct the vector $\bx = [0,1,2,\ldots,N]/N$.\\
\> \> Construct the vector $\bw = [1,4,2,4,2,\ldots,4,2,1]$ as in \eqref{eq:wvec} \\
\> \> Compute the path length $p_i$ from the root to each node $i$.\\
\> \> Initialize $F_i[k] \leftarrow 1$ for all nodes $i$ and $0 \leq k \leq N$.\\
\> \> For all $i = n+1,n+2,\ldots,2n\!-\!1$ \\
\> \> \>  $L_i \leftarrow \Phi^{-1}(\frac{nN^{-4}}{2(n-1)}|\mu_r,\sigma_r^2 + \sigma^2 p_i)$\\
\> \> \>  $U_i \leftarrow \Phi^{-1}(1-\frac{nN^{-4}}{2(n-1)}|\mu_r,\sigma_r^2 + \sigma^2 p_i)$\\
\> \> \> $X_i \leftarrow (U_i - L_i) \bx + L_i$\\
\> \> For all tip nodes  $i = 1,2,\ldots,n$\\
\> \> \> Let $j$ be the index of the parent of node $i$\\
\> \> \> For $k=0,\ldots,N$\\
\> \> \> \> If $z_i = 1$\\
\> \> \> \> \> $F_i[k] = 1-\Phi(0;X_j[k],\sigma^2 t_i)$ \\
\> \> \> \> else if $z_i = 0$\\
\> \> \> \> \> $F_i[k] =\Phi(0;X_j[k],\sigma^2 t_i)$ \\
\> \> For all internal nodes $i = n\!+\!1,...,2n\!-\!2$, excluding the root\\
\> \> \>  Let $j$ be the index of the parent of node $i$ \\
\> \> \> Let $u,v$ be the indices of the children of node $i$\\
\> \> \> For $k=0,1,\ldots,N$\\
\> \> \> \>   $\displaystyle F_i[k] \leftarrow \frac{U_i-L_i}{N} \sum \displaylimits_{\ell = 0}^N \bw_\ell \phi(X_i[\ell]; X_j[k],\sigma^2 t_i) F_u[\ell] F_v[\ell]$\\
\> \> Let $u,v$  be indices of the the children of the root.\\
\> \> $\displaystyle L \leftarrow \frac{U_{2n-1} - L_{n-1}}{N} \sum_{\ell = 0}^N \bw_\ell \phi(X_i[\ell]; \mu_r,\sigma_r^2) F_u[\ell] F_v[\ell]$
\end{tabbing}
\end{sffamily}
\caption{\label{algo:OneChar} Pseudo-code of the likelihood approximation algorithm for a single character, under the threshold model. The nodes are numbered in increasing order from tips to the root.}
\end{algorithm}


We can estimate posterior densities using the recursion \eqref{eq:postGeneral} followed by equation \eqref{eq:posteq}. The conditional density 
\begin{equation} f(\tilde{x}|x,\theta_i) = \phi\left(\tilde{x} \Big| \mu_r + \frac{\sigma_r^2 + \sigma^2 P_j}{\sigma_r^2 + \sigma^2 P_i} \left( x - \mu_r \right)  , \frac{\sigma^2 t_i \left( \sigma_r^2 + \sigma^2 P_j \right)}{\sigma_r^2 + \sigma^2 P_i} \right)\end{equation} 
can be obtained by plugging the transitional density
\begin{equation} f(x| \tilde{x},\theta_i) = \phi(x|\tilde{x},\sigma^2 t_i)\end{equation} 
and the two marginal densities  \eqref{eq:marginalv}
\begin{equation} f(\tilde{x}) = \phi(\tilde{x},\sigma_r^2 + \sigma^2 P_j),\quad f(x) = \phi(x,\sigma_r^2 + \sigma^2 P_i)\end{equation} 
into the identity
$f(\tilde{x}|x,\theta_i) = f(x| \tilde{x},\theta_i)  \frac{f(\tilde{x})}{f(x)}$. We thereby obtain the recursion 
\begin{equation}
\sG_i(x) = \int \phi\left(\tilde{x} \Big| \mu_r + \frac{\sigma_r^2 + \sigma^2 P_j}{\sigma_r^2 + \sigma^2 P_i} \left( x - \mu_r \right)  , \frac{\sigma^2 t_i \left( \sigma_r^2 + \sigma^2 P_j \right)}{\sigma_r^2 + \sigma^2 P_i} \right) \sG_j(\tilde{x}) \sF_u(\tilde{x}) \, d \tilde{x}\label{eq:postThreshold}
\end{equation}
which we estimate using Simpson's method. Algorithm estimates values of the  posterior densities 
at each node, evaluated using the same set of grid points as used in Algorithm 1. An additional round of numerical integration can be used to obtain posterior means and variances.

\begin{algorithm}[htb]
\begin{sffamily}
\begin{tabbing}
\=XXX\=XXX\=XXX\=XXX\=XXX\=XXX\= \kill
{\bf Algorithm 2}: Compute posterior probabilities\\ ~\\
\> \> {\bf Input:} \\
\> \> \> $N$, $t_1,\ldots 2n-2$, $\mu_r$, $\sigma^2_r$, and $\sigma^2$ as in Algorithm 1\\
\> \> \> Vector $p$, likelihood $L$ and arrays $F_i$ computed in Algorithm 1.\\
\> \> {\bf Output:}\\
\> \> \> Arrays $P_i$ for each internal node $i$.
\\
\> \> Construct the vectors $\bx$, $\bw$, $L$, $U$, and path lengths $p_i$ as in Algorithm 1.\\
\> \> $G_{2n-1}[k] \leftarrow 1$ for all $k$.\\
\> \> For all $i = 2n\!-\!2,2n-2,\ldots,n+1$\\
\> \> \> Let $j$ be the index of the parent of node $i$.\\
\> \> \> Let $v$ be the index of the sibling of node $i$.\\
\> \> \> For $k=0,1,\ldots,N$\\
\> \> \> \> $\mu \leftarrow \mu_r + \frac{\sigma_r^2 + \sigma^2 P_j}{\sigma_r^2 + \sigma^2 P_i} \left( X_i[k] - \mu_r \right)$\\
\> \> \> \> $V \leftarrow \frac{\sigma^2 t_i \left( \sigma_r^2 + \sigma^2 P_j \right)}{\sigma_r^2 + \sigma^2 P_i}$\\
\> \> \> \> $\displaystyle G_i[k] \leftarrow  \frac{U_j-L_j}{N} \sum \displaylimits_{\ell = 0}^N \bw_\ell \phi(X_j[\ell]; \mu,V) G_j[\ell] F_v[\ell]$\\
\> \> For all $i = n+1,\ldots,2n-1$\\
\> \> \> Let $u,v$ be the children of node $i$.\\
\> \> \> For all $k = 0,1,\ldots,N$\\
\> \> \> \> $P_i[k] \leftarrow \frac{1}{L} G_i[k] F_u[k] F_v[k] \phi(X_i[k]|\mu_r,\sigma_r^2 + \sigma^2 p_i)$\\
\end{tabbing}
\end{sffamily}
\caption{\label{algo:Posterior} Pseudo-code for the algorithm to efficiently compute ancestral posterior  densities under the threshold model. At the termination of the algorithm, $P_i[k]$ is an estimate of the posterior density at internal node $i$, evaluated at $x = X_i[k]$.}
\end{algorithm}

\subsubsection*{Evolutionary precursors of plant extrafloral nectaries}

To study the methods in practice, we reanalyse trait data published by \cite{MarazziETAL2012a}, using a fixed phylogeny.
\cite{MarazziETAL2012a} introduce and apply a new discrete state model for morphological traits which, in addition to states for presence and absence, incorporates an intermediate `pre-cursor' state. Whenever the intermediate state is observed at the tips it is coded as `absent'. The motivation behind the model is that the intermediate state represents evoutionary pre-cursors, changes which are necessary for the evolution of a new state but which may not be directly observed. These pre-cursors could explain repeated parallel evolution of a trait in closely related traits \citep{MarazziETAL2012a}. They compiled a data set recording presence or absence of plant extrafloral nectaries (EFNs) across a phylogeny of 839 species of Fabales, fitting their models to these data. 

The threshold model also involves evolutionary pre-cursors in terms of changes in ancestral liabilities. We use these models, and our new algorithms to analyse the EFN dataset. Our analysis also makes use of the time-calibrated phylogeny inferred by \cite{SimonETAL2009a}, although unlike \cite{MarazziETAL2012a} we ignore phylogenetic uncertainty.

\subsubsection*{Experimental protocol}

We conduct three separate experiments. For the first experiment, we examine the rate of convergence of the likelihood algorithm as we increase $N$. This is done for the `All'  EFN character (Character 1 in  \cite{MarazziETAL2012a}) for a range of estimates for the liability variance at the root, $\sigma_r^2$. The interest in $\sigma_r^2$ stems from its use in determining bounds $L_i,U_i$ for each node, with the expectation that as $\sigma_r^2$ increases, the convergence of the integration algorithm will slow. The mean liability at the root, $\mu_r$ was determined from the data using Maximum Likelihood estimation.

We also examined convergence of the algorithm on randomly generated characters. We first evolved liabilities according to the threshold model, using the parameter settings obtained above. To examine the difference in performance for {\em non-phylogenetic} characters we also simulated binary characters by simulated coin flipping. Twenty replicates were carried out for each case.

The second experiment extends the model comparisons carried out in \cite{MarazziETAL2012a} to include the threshold models. For this comparison we fix the transitional variance $\sigma^2$ at one, since changing this values corresponds to a rescaling of the Brownian process, with no change in likelihood. With only one character, the maximum likelihood estimate of the root variance $\sigma_r^2$ is zero, irrespective of the data. This leaves a single parameter to infer: the value of the liability at the root state. We computed a maximum likelihood estimate for the state at the root, then applied our algorithm with a sufficiently large value of $N$ to be sure of convergence. The Akaike Information Criterion (AIC) was determined and compared with those obtained for the model of  \cite{MarazziETAL2012a}.

For the third experiment, we determine the marginal posterior densities for the liabilities at internal nodes, using Algorithm~\ref{algo:Posterior}. These posterior probabilities are then mapped onto the phylogeny, using shading to denote the (marginal) posterior probability that a liability is larger than zero. We therefore obtain a figure analogous to Supplementary Figure 7 of \cite{MarazziETAL2012a}.

\section*{Results}

\subsection*{Convergence of the algorithm}

\begin{figure}[htb]
\centerline{\includegraphics[width=1.2\textwidth]{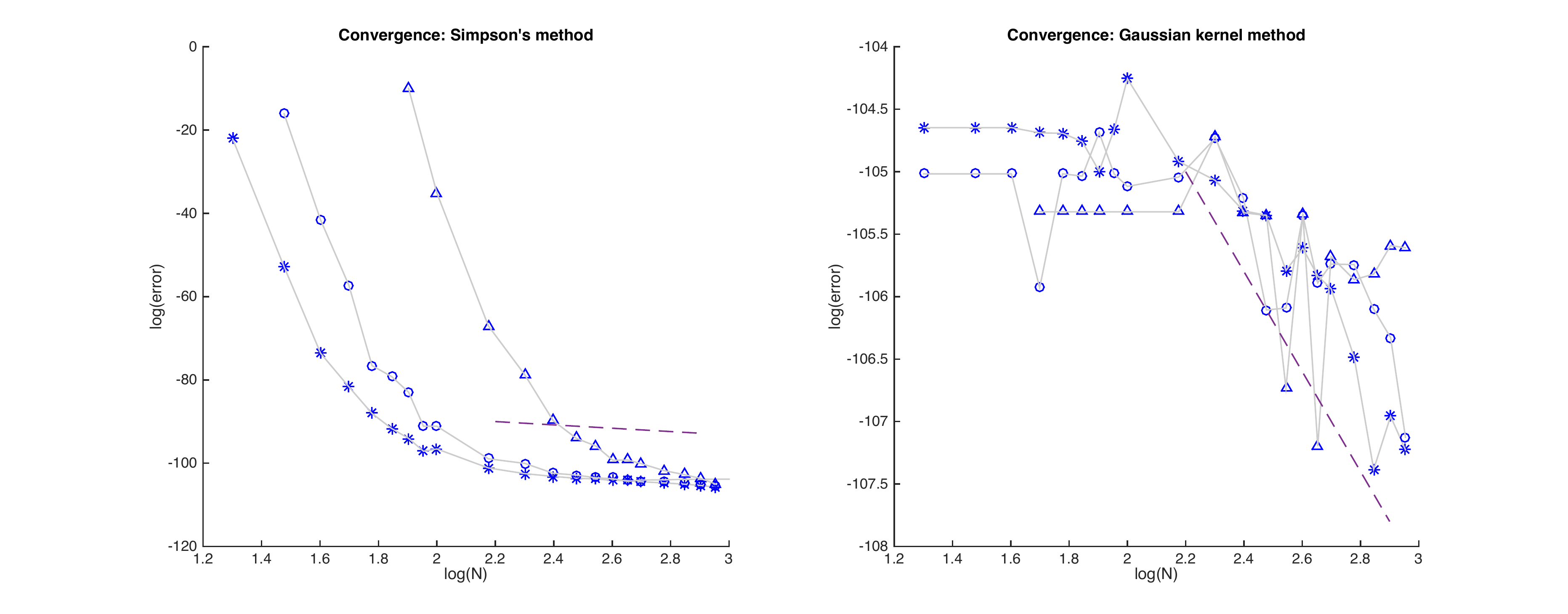}}
\caption{\label{fig:converge1} Log-log plots of error as a function of $N$ for the dynamic programming algorithm with Simpson's method (left) and with the Gaussian kernel method (right). The likelihoods were computed under the threshold model on EFN trait data for an 839 taxon tree. Dotted lines have slope -4 (corresponding to convergence rate of $N^{-4}$. Note the difference in scale for the two methods.). Logarithms computed to base 10.}
\end{figure}

\begin{figure}[htb]
\centerline{\includegraphics[width=\textwidth]{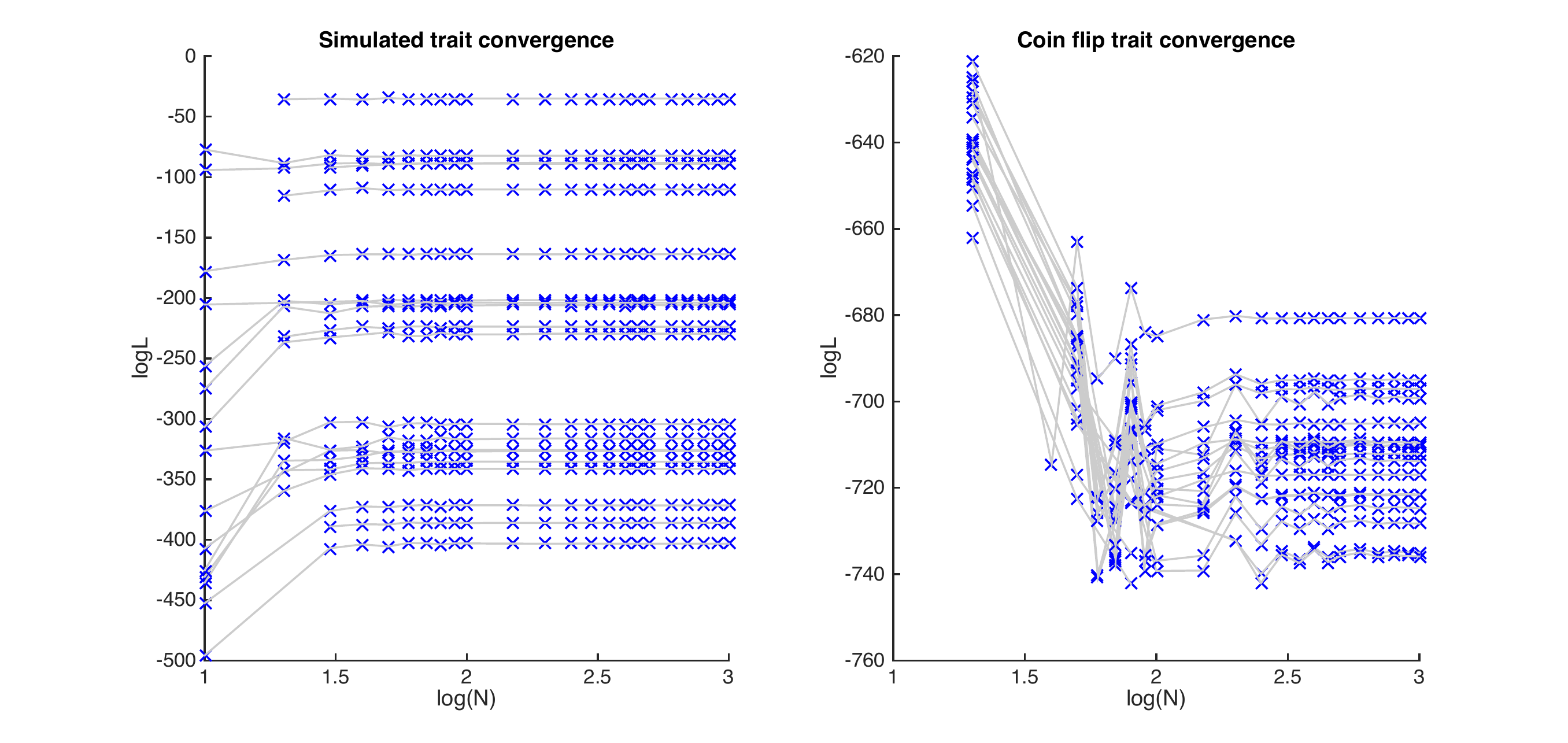}}
\caption{\label{fig:converge2} Plots of log-likelihood values as a function of $\log(N)$ for the two types of simulated data, computed using our algorithm together with the Gaussian kernel method. Logarithms computed to base 10.}
\end{figure}

Plots of error versus $N$ are given in Figure~\ref{fig:converge1}, both for Simpson's method  (left) and for the modified Gaussian kernel method (right). For larger $N$, the error in a log-log plot decreases with slope at most $-4$ (as indicated), corresponding to $N^{-4}$ convergence of the method. Log-log plots of error versus $N$ for the simulated data are given in Figure~\ref{fig:converge2}. In each case, the method converges for by $N \approx 30$.

While the level of convergence for both algorithms is correct, the accuracy of the method based on Simpson's method is far worse. When a branch length is short, the transition density becomes highly peaked, as does the function being integrated. Such functions are difficult to approximate with piecewise quadratics, and Simpson's method can fail miserably. Indeed, for $N < 50$, we would often observe negative estimated probabilities, or estimates greater than one! (These were omitted from the plots). While we can always bound estimates computed by the algorithm, a sounder approach is to improve the integration technique. This we did using the Gaussian kernel method, and the result was far improved accuracy for little additional computation. For the remainder of the experiments with this model we used the Gaussian kernel method when carrying out numerical integration. 

\subsection*{Model comparison}

\cite{MarazziETAL2012a} describe AIC comparisons between their pre-cursor model and a conventional binary trait model. We extend this comparison to include the threshold model. This is a one parameter model, the parameter being the value of the liability at the root. We used the MATLAB command {\tt fminsearch} with multiple starting points to compute the maximum likelihood estimate for this value. The resulting log-likelihood was $\log L = -240.6$, giving an AIC of $483.2$. This compares to an AIC of $507.4$ for the (two parameter) binary character model and an AIC of $495.4$ for the (one parameter) precursor model of \cite{MarazziETAL2012a}.

We analyzed the five other EFN traits in the same way, and present the computed AIC values in Table~\ref{tab:aic}, together with AIC values for the two parameter binary state model and one parameter precursor model computed by 
\cite{MarazziETAL2012a} (and the 2 parameter precursor model for trait 6). We see that the threshold model fits better than either the binary or precursor models for all of  the six traits. 

\begin{table}[htb]
\begin{center}
\begin{tabular}{llccc}
\hline \hline
Trait & Model & $k$ & $\log L$ & AIC \\
\hline
1 (All) & Binary & 2 & -251.7 & 507.4 \\
  & Precursor & 1 & -246.7 & 495.4 \\
 & Threshold & 1 & -240.6 & {\bf 483.2}  \\
2 (Leaves) & Binary & 2 & -240.3 & 484.6 \\
  & Precursor & 1 & -234.5 & 470.9\\
 & Threshold & 1 & -230.6 &  {\bf 463.1}\\
3 (Inflorescence) & Binary & 2 & -108.3 & { 220.5} \\
  & Precursor & 1 & -110.9 & 223.9 \\
 & Threshold & 1 & -108.3 & {\bf 218.5} \\
4 (Trichomes) & Binary & 2 & -86.7 & 177.3\\
  & Precursor & 1 & -86.9 & 325.3 \\
 & Threshold & 1 & -85.8 &  {\bf 173.5}\\
5 (Substitutive) & Binary & 2 & -163.0 & 330.1 \\
  & Precursor & 1 & -161.6 & 325.3 \\
 & Threshold & 1 & -161.3 & {\bf 324.6} \\
6 (True) & Binary & 2 & -132.3.1 & 268.7 \\
  & Precursor & 1 & -131.1 & 264.3 \\
& Precursor & 2 & -126.7 & { 257.3} \\
 & Threshold & 1 & -125.3 & {\bf 252.6} \\
\hline
\end{tabular}
\end{center}
\caption{\label{tab:aic} Table of log-likelihood and AIC values for the binary character, precursor, and threshold models on six EFN traits. Column $k$ indicates numbers of parameters for each model. Data for the binary and precursor models copied from Table 1 in \cite{MarazziETAL2012a}. All likelihoods and AIC values rounded to 1 d.p. Boldface indicates the best fitting model for each trait. }
\end{table}

It is not clear, {\em a priori}, why the threshold model would appear to fit some data  better than the precursor model since they appear to capture similar evolutionary phenomena. It would be useful to explore this observation more thoroughly, given the new computational tools, perhaps incorporating phylogenetic error in a manner similar to  \cite{MarazziETAL2012a}.

\subsection*{Inferring ancestral liabilities}

\begin{figure}
\centerline{\includegraphics[width=\textwidth]{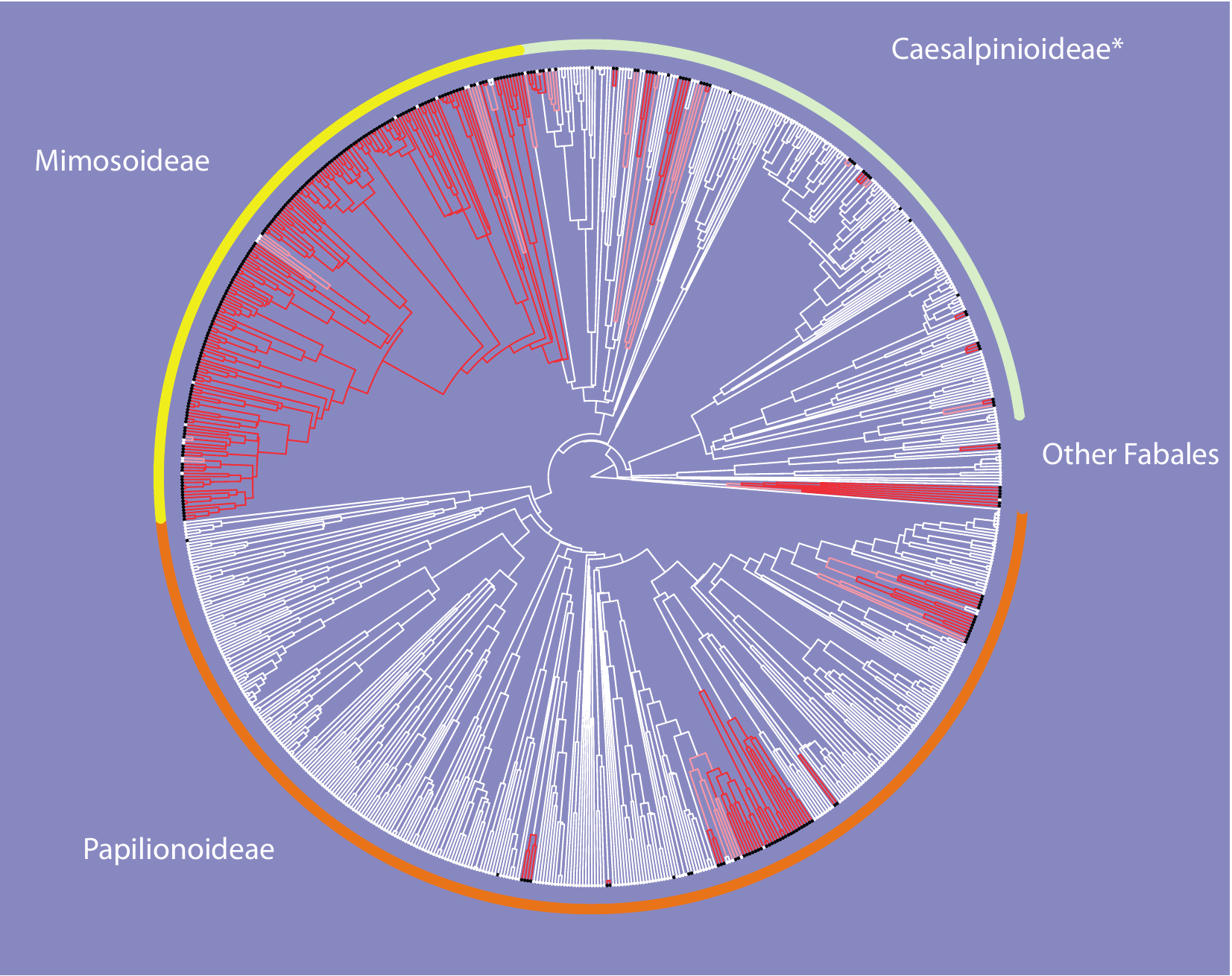}}
\caption{\label{fig:prettyTree} Marginal posterior probabilities for the liabilities, for EFN trait 1 of \cite{MarazziETAL2012a} on the phylogeny inferred by \cite{SimonETAL2009a}. Lineages with posterior probability $> 0.7$ colored red, lineages with posterior probability $< 0.3$ colored white, and remaining lineages colored pink.  
}
\end{figure}

Figure~\ref{fig:prettyTree} gives a representation of how the (marginal) posterior liabilities change over the tree. Branches are divided into three classes according to the posterior probability that the liability is positive, with lineages with posterior probability $> 0.7$ colored red, lineages with posterior probability $< 0.3$ colored white, and remaining lineages colored pink. 

This diagram can be compared to Supplementary Figure~7, of \cite{MarazziETAL2012a}. The representations are, on the whole, directly comparable.  An positive liability corresponds, roughly, to an ancestral precursor state. Both analyses suggest multiple origins of a precursor state, for example for a large clade of Mimosoidae. Interestingly, there are several clades where the analysis of \cite{MarazziETAL2012a} suggests widespread ancestral distribution of the precursor state whereas our analysis indicates a negative liability at the same nodes.

Once again, our analysis is only preliminary, our goal here simply being to demonstrate what calculations can now be carried out. 

\section*{Discussion}

We have introduced a new framework for the computation of likelihoods from continuous characters, and illustrated the framework using an efficient algorithm for evaluating (approximate) likelihoods under Wright and Felsenstein's threshold model. 

This framework opens up possibilities in several directions. The numerical integration, or numerical quadrature, literature is vast. In this article, we have focused in on a popular and simple numerical integration method, and our algorithm should be seen as a proof of principle rather than a definitive threshold likelihood method. There is no question that the numerical efficiency of Algorithm 1 could be improved significantly through the use of more sophisticated techniques: better basis functions or adaptive quadrature methods for a start. 

The connection with Felsenstein's (discrete character) pruning algorithm also opens up opportunities for efficiency gains. Techniques such as storing partial likelihoods, or approximating local neighborhoods, are fundamental to efficient phylogenetic computations on sequence data \citep{Felsenstein1981b,LargetETAL1998a,PondETAL2004a,Stamatakis2006a,Swofford2003a}. These tricks could all be now applied to the calculation of likelihoods from continuous traits.

Finally, we stress that the algorithm does not depend on special characteristics of the continuous trait model, beyond conditional independence of separate lineages. Felsenstein's pruning algorithm for continuous characters is limited to Gaussian processes and breaks down if, for example, the transition probabilities are governed by Levy processes \citep{LandisETAL2013a}. In contrast, our approach works whenever we can numerically evaluation transition densities, an indeed only a few minor changes would transform our Algorithm 1 to one implementing on a far more complex evolutionary process.

\section*{Acknowledgements}

This research was supported by an Allan Wilson Centre Doctoral Scholarship to GH, financial support to DB from the Allan Wilson Centre, a Marsden grant to DB, and financial support to all authors from the University of Otago.

\bibliographystyle{plain}  
\bibliography{HiscottEtalGBE}

\end{document}